# Field-induced multiple metamagnetization in phase transition from paramagnetic austenite to ferromagnetic martensite in $MnNi_{1-x}Fe_xGe$

E.K. Liu, G.J. Li, W. Zhu, L.Feng, J. L. Chen, G. H. Wu, and W. H. Wang*

*Beijing National Laboratory for Condensed Matter Physics, Institute of Physics, Chinese Academy of Sciences, Beijing 100190, China*

**Abstract** - Tailoring the phase stability and magnetic structure by combining the antiferromagnetic MnNiGe martensite with ferromagnetic MnFeGe austenite, a magnetic-field-induced martensitic transformation has been achieved in $MnNi_{1-x}Fe_xGe$ system. The doped Fe enables the isolated structural and magnetic transitions of MnNiGe to coincide at room temperature and also establishes ferromagnetic couplings in six-Mn-centering-Fe-atom clusters. The ferromagnetic clusters transit the antiferromagnetic state in martensite to a ferromagnetic state and the alloys are further induced to a stronger ferromagnetic state by external field, resulting in a large magnetization difference up to 50 emu/g across the transformation. The field-induced martensitic transformation thus experiences a multiple metamagnetization process.

* E-mail: wenhong.wang@iphy.ac.cn

Magnetic martensitic transformation has become more attractive in many systems for magnetic-field-induced shape memory alloys,[1-4] magnetoresistance[5, 6] and magnetic refrigerations.[7-9] Across the transformation, a large magnetization difference ($\Delta M$) between two phases is widely deemed to be of great importance for these multifunctional applications. In some magnetic equiatomic MM'X (M, M' = transition metals, X = Si, Ge, Sn) compounds, there exist martensitic transformations between high-temperature hexagonal $Ni_2In$-type structure and low-temperature orthorhombic TiNiSi-type one.[10, 11] However, the martensitic transformation almost occurs in the paramagnetic (PM) states, in which an appreciable magnetic transition is absent.[10] Recently, it has been found that the transition-metal vacancies and doping in these compounds can remarkably lower the martensitic-transformation temperatures down below their Curie temperatures ($T_C$).[12-15] With a resultant large $\Delta M$, magnetic-field-induced martensitic transformations and associated magnetocaloric effects have been increasingly obtained in off-stoichiometric MnCoGe and MnNiGe systems.[14, 16-18]

As an attractive candidate, stoichiometric MnNiGe transforms to an antiferromagnetic (AFM) martensite at 470 K (in PM state).[19] It can thus not exhibit an appreciable $\Delta M$ for magnetic martensitic transformation. Another isostructural compound of MnFeGe, contrastively, has a stable ferromagnetic (FM) $Ni_2In$-type phase down to 4.2 K.[20] In these two systems, Ni and Fe atoms occupy the same atomic sites.[20] So it seems to be possible to obtain a quaternary system of MnNiFeGe in which the martensitic transformation may be lowered to a proper temperature, where the martensitic and a PM-to-FM transitions can coincide with a generated large $\Delta M$ for a magnetic-field-controllable martensitic transformation.

In this letter, we report that this desired PM/FM martensitic transformation have been realized in Fe-doped MnNiGe system around room temperature. Due to the large *ΔM*, the field-induced martensitic transformation becomes available. It has been found that the strong exchange interaction established by doped Fe and the Zeeman energy provided by the external field together turn the AFM magnetic structure to a FM one. The magnetic-field-induced martensitic transformation thus undergoes a special multiple metamagnetic process in which the magnetic and structural transitions also have the same sign in thermodynamic exothermic/endothermic behaviors.

Polycrystalline ingots of $MnNi_{1-x}Fe_xGe$ were prepared by arc-melting raw metals in high-purity argon atmosphere. The ingots were annealed in evacuated quartz tube with Ar gas at 1123 K for five days and cooled with furnace to room temperature. The room-temperature structures were identified by powder x-ray diffraction (XRD) with Cu *Kα* radiation. Magnetization measurements were carried out using a superconducting quantum interference device (SQUID) on the powder samples. The differential thermal analysis (DTA) method with heating and cooling rate of 2.5 K/min is also used to measure the martensitic transformation data.

Figure 1 shows the room-temperature XRD patterns for $MnNi_{1-x}Fe_xGe$ ($x$=0, 0.20, 0.23 and 0.27). The Fe-free MnNiGe shows an orthorhombic martensite phase at room temperature and the hexagonal parent phase gradually appears with increasing Fe content. This indicates that the martensitic transformations can indeed be lowered down by doping Fe. For $MnNi_{0.80}Fe_{0.20}Ge$, the lattice constants of austenite and martensite are determined as $a_h$ = 4.0825 μm, $c_h$ = 5.4029 μm and $a_o$ = 6.0260 μm, $b_o$ = 3.7763 μm, $c_o$ = 7.0959 μm, respectively. The unit-cell volume increases by about 3.53% during the transformation, which is larger than that of stoichiometric MnNiGe (1.6%).[19]

Figure 2 shows the *M(T)* curves of $MnNi_{1-x}Fe_xGe$ in a field of 50 kOe. A quite narrow thermal hysteresis about 10 K in the martensitic transformation is clearly presented. The martensitic-transformation starting temperatures, $T_m$ and other related parameters are measured by the various methods as listed in Table I. For the low level doping of just only $x = 0.20$, the $T_m$ of the sample is already at 303 K (measured in low field), much lower than the $T_m$ (470 K) of stoichiometric MnNiGe and $T_N$ (346 K) of the stoichiometric martensite.[19] However, the curve shows a strong FM response rather than an AFM one, which indicates that the sample evidently undergoes a transformation from a PM austenite to a FM martensite. For higher doping, the other two alloys have the same transformation behavior, as also seen in Table I. The saturation magnetization at 5 K reaches about 100 emu/g, associated with a large $\Delta M$ of about 50 emu/g [also see Table I]. For $MnNi_{0.77}Fe_{0.23}Ge$, the $T_m$ shifted up about 11 K by the field of 50 kOe relative to that observed in the low field, as shown in the inset of Fig. 2. This situation is quite similar to that of $Fe_2MnGa$ and MnCoGe alloys,[16, 21, 22] in which a PM/FM-type martensitic transformation occurs with a large $\Delta M$, enabling a magnetic-field-induced martensitic transformation. Accordingly, one can see that the AFM character in the martensite of MnNiGe conspicuously transits to a FM one in $MnNi_{1-x}Fe_xGe$ system.

To trace the AFM/FM transition nature, we explore the magnetization behaviors of $MnNi_{1-x}Fe_xGe$ martensites up to 50 kOe at 5 K, as shown in Fig. 3. The Fe-free sample shows a typical antiferromagnetic behavior with a kink point exhibiting a metamagnetic transition from an AFM to a FM state, as reported in previous work,[19] in which a very high field up to about 100 kOe can make the system ferromagnetically saturate. In sharp contrast, the Fe-doped martensite strikingly show a large and increasing initial $dM/dH$ and the kink point (critical field, $H_{cr}$) rapidly decreases from 11.5 to 1.7 kOe, leaving the alloy with highest doping-level ($x$=0.27) an almost-complete FM ground state with the magnetization up to about 100 emu/g, about twice as big as that of the Fe-free sample. The saturated magnetization and $H_{cr}$ of all samples are illustrated in the inset and Table I.

Based on the crystal structure of MnNiGe system,[19] the AFM exchange interaction exists between Mn atoms due to their large separation and the zero moment of Ni atoms, as depicted in Fig. 4(a). Once a Ni atom is replaced by a Fe atom with non-zero magnetic moment, short term strong FM exchange interactions largely establish in a cluster with six neighbor Mn centering a Fe atom, which is similar to the case for doping Co in this system.[23] Thus the AFM cluster becomes a Fe-6Mn FM one, as shown in Fig. 4(b). With increasing Fe doping, the FM clusters increase in quantity. In case of ideal uniformity, therefore, $x$ = 0.25 becomes a critical composition at which each unit cell possesses one FM cluster. With higher doping, the subsequent Fe atoms will further ferromagnetically couple two FM clusters in neighbor unit cells to eventually complete the AFM/FM transition in the martensite. In this process, the competition between AFM and FM orderings continuously exists and the applied magnetic field thus further drive the FM state to a stronger FM one as the introduced Zeeman energy favors the strong FM state. The effect of magnetic field starts from the kink point of each curve as shown in Fig. 3. The spiral AFM structure thus changes to a parallel structure in FM state by combined FM exchange interactions provided by both the doped Fe and the magnetic field. Two of them work together to promote the AFM/FM transition, resulting in the high magnetization of $MnNi_{1-x}Fe_xGe$ martensite.

Fig. 5 shows the magnetization of $MnNi_{0.77}Fe_{0.23}Ge$ at various temperatures on cooling across the martensitic transformation. Above 279 K, the austenite shows a PM behavior. Between 274 and 250 K, the magnetization curves notably jump at about $H_{cr}^{MT}$ = 20 kOe, showing a new metamagnetic phase transition with an apparent hysteresis. This metamagnetization corresponds to the magnetic-field-induced martensitic transformation from PM austenite to FM martensite which has been expected when we observed the results shown in Fig. 2. The distinct $\Delta M$ between the martensite and PM austenite, shown in Fig. 3 and Table I, here introduces a larger Zeeman energy for high-magnetization martensite in applied field, giving rise to an energetically favored martensite. At about $H_{cr}$ = 3 kOe, it can be seen that the produced martensite experiences a prior FM/FM magnetic transition, which has been discussed in Figs. 3 and 4. One can thus see that, as a consequence of the AFM/FM transition in $MnNi_{1-x}Fe_xGe$ system, the FM martensite lays the first stone for the subsequent field-induced PM/FM martensitic transformation. Contrastively, in Co-doped MnNiGe,[23] although the AFM state of martensite has completely transited to a FM state due to the Co doping, the martensitic transformation still occurs above the $T_C$ and, the magnetic and structural transitions cannot coincide. It deserves to further note that, for this PM/FM-type martensitic transformation in $MnNi_{1-x}Fe_xGe$ system, the coincident PM/FM magnetic and structural transitions would show the same sign of exothermic/endothermic behaviors on cooling and heating, which enhances the caloric effect for practical applications.

In summary, combining the AFM MnNiGe martensite with FM MnFeGe austenite to tailor the phase stability and magnetic structure, an AFM/FM transition and martensitic transformation has been integrately achieved in MnNiFeGe system around the room temperature with a large $\Delta M$ of about 50 emu/g. It has been found that the doped Fe and six neighbour Mn atoms form Fe-6Mn FM clusters to enhance the ferromagnetic exchange interaction and make the AFM changed to a FM state which further transits to a stronger FM state by the applied field. Thus the magnetic-field-induced martensitic transformation experiences a multiple metamagnetization process. These multiple phase transitions in MnNiFeGe system may make it potentially interesting as multifunctional materials, such as field-induced strain and caloric effect at room temperature.

This work is supported by the National Natural Science Foundation of China in Grant No. 50771103 and 50971130.

# FIGURE CAPTIONS

Table I. Values of $T_{\mathrm{m}}$ , $\Delta T$, $M$ , $\Delta M$ and $H_{\mathrm{cr}}$ for $\mathrm{MnNi}_{1\text{-}x}\mathrm{Fe}_x\mathrm{Ge}$ alloys.

Fig. 1: (Color online) Power XRD patterns of $\mathrm{MnNi}_{1\text{-}x}\mathrm{Fe}_x\mathrm{Ge}$ measured at room temperature. The dashed lines denote the positions of Bragg peaks of hexagonal austenite phase.

Fig. 2: (Color online) Temperature dependence of the magnetization of $\mathrm{MnNi}_{1\text{-}x}\mathrm{Fe}_x\mathrm{Ge}$ in the field of 50 kOe. The $\Delta M$ denotes a magnetization difference between the martensite and austenite phases. The inset shows the magnetization curves for the selected $\mathrm{MnNi}_{0.77}\mathrm{Fe}_{0.23}\mathrm{Ge}$ and the magnetization in the field of 500 Oe is enlarged by a factor of 10.

Fig. 3: (Color online) Magnetization isotherms at 5 K of $\mathrm{MnNi}_{1\text{-}x}\mathrm{Fe}_x\mathrm{Ge}$ martensites in fields up to 50 kOe. The inset shows Fe-doping dependences of the critical field ($H_{\mathrm{cr}}$) and magnetization at 5 K.

Fig. 4: Schematic of the AFM structure in MnNiGe martensite (a) and Fe-6Mn FM cluster in $\mathrm{MnNi}_{1\text{-}x}\mathrm{Fe}_x\mathrm{Ge}$ martensite (b).

Fig. 5: (Color online) Magnetization isotherms of $\mathrm{MnNi}_{0.77}\mathrm{Fe}_{0.23}\mathrm{Ge}$ at various temperatures across the martensitic transformation.

| $x$ | $T_m$ | $\Delta T$ | $M$ | $\Delta M$ | $H_{cr}$ |
|---|---|---|---|---|---|
| | K | | emu/g | | kOe |
| 0 | 483 | 26 | 48 | - | 11.5 |
| 0.20 | 303 | 14 | 91 | 46 | 5.4 |
| 0.23 | 273 | 11 | 97 | 51 | 4.3 |
| 0.27 | 236 | 10 | 100 | 49 | 1.7 |

# Table I

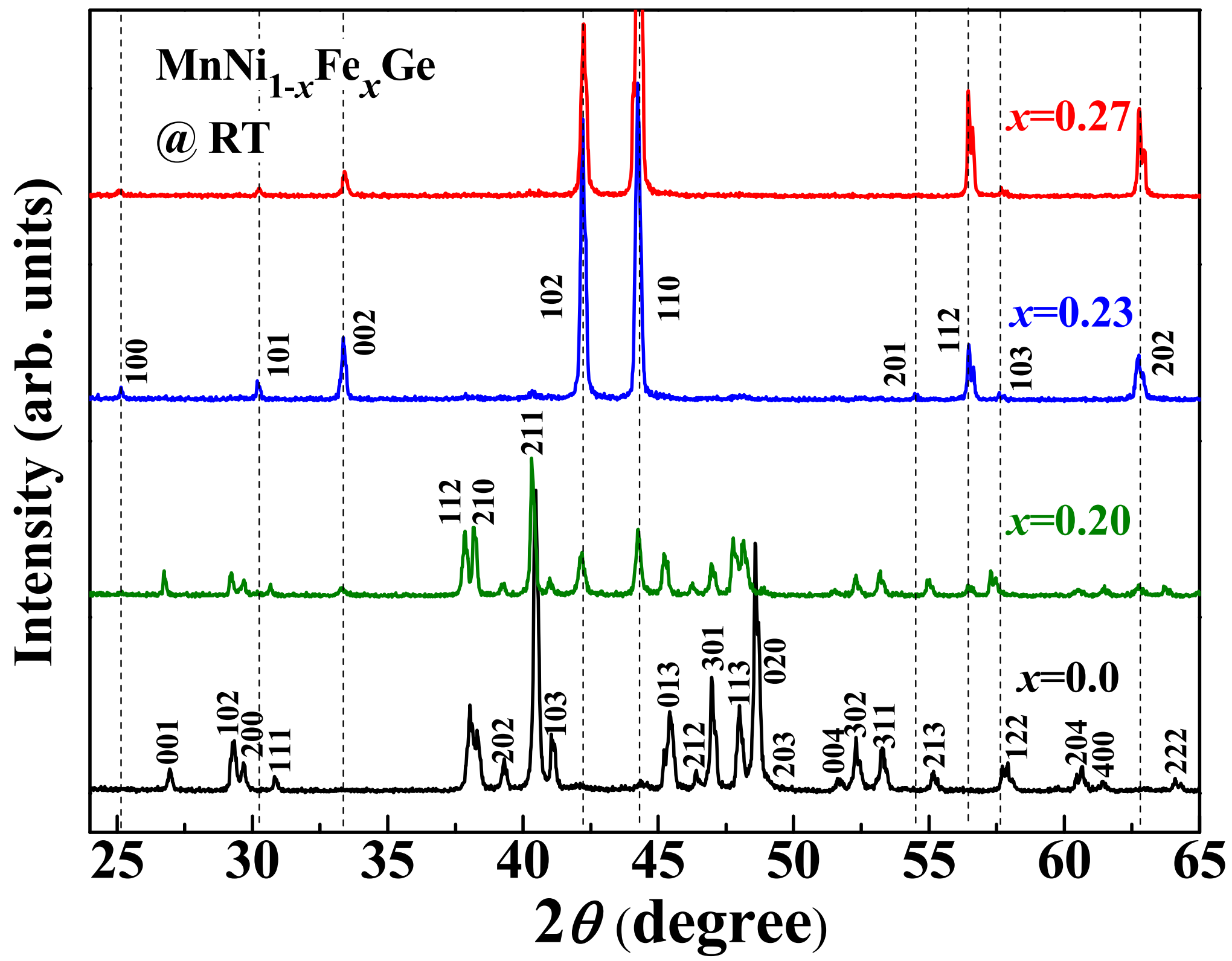

Fig. 1

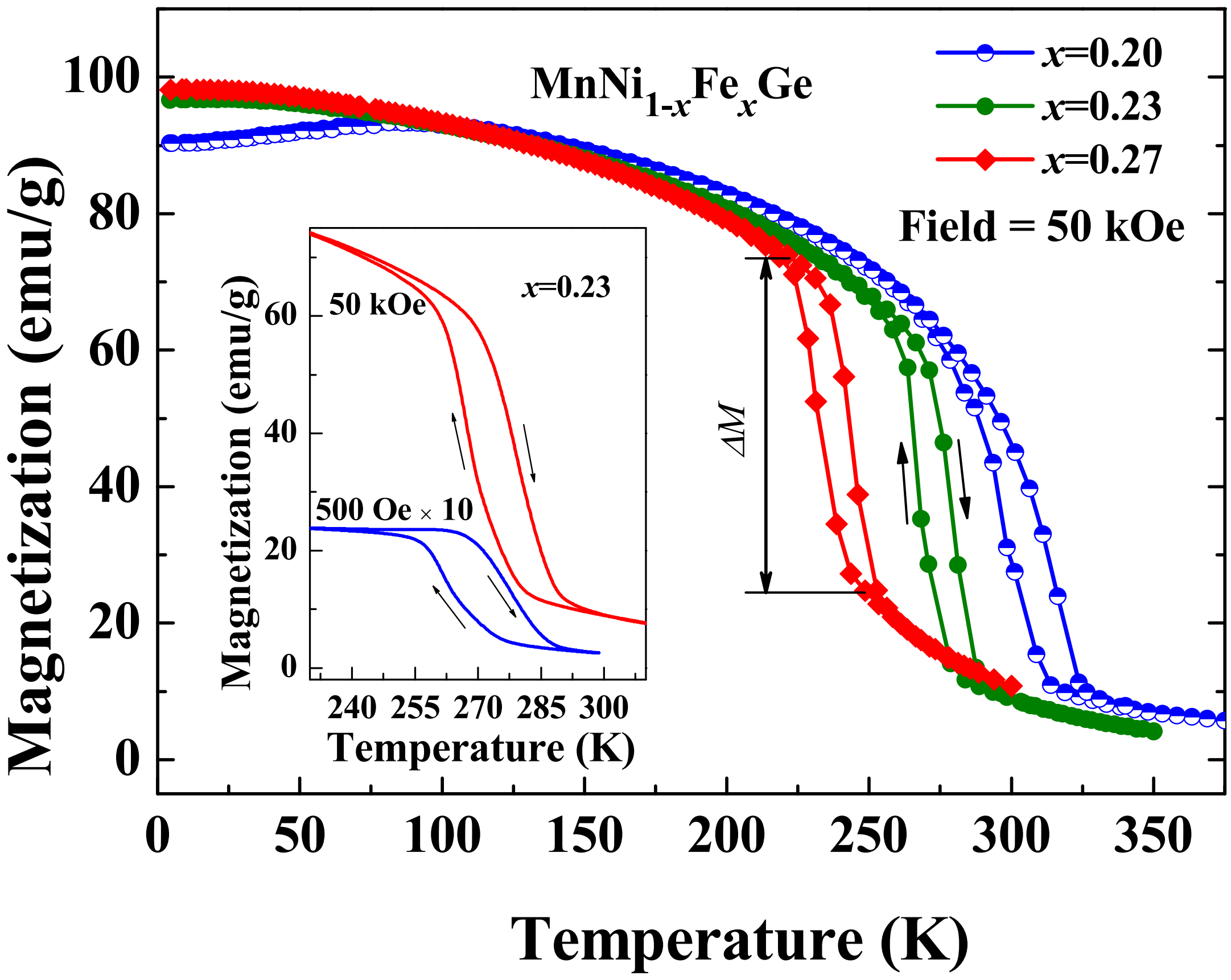

Fig. 2

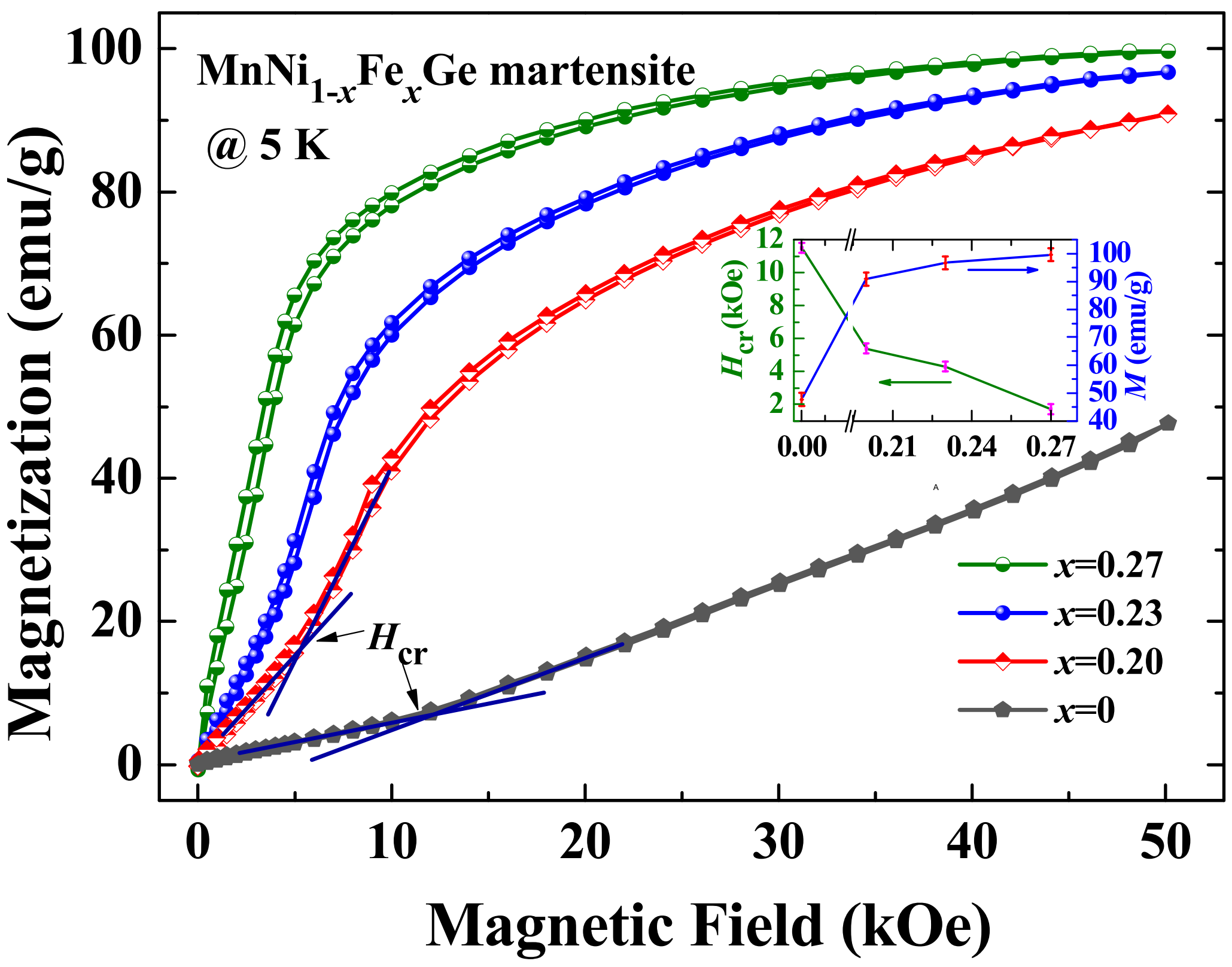

Fig. 3

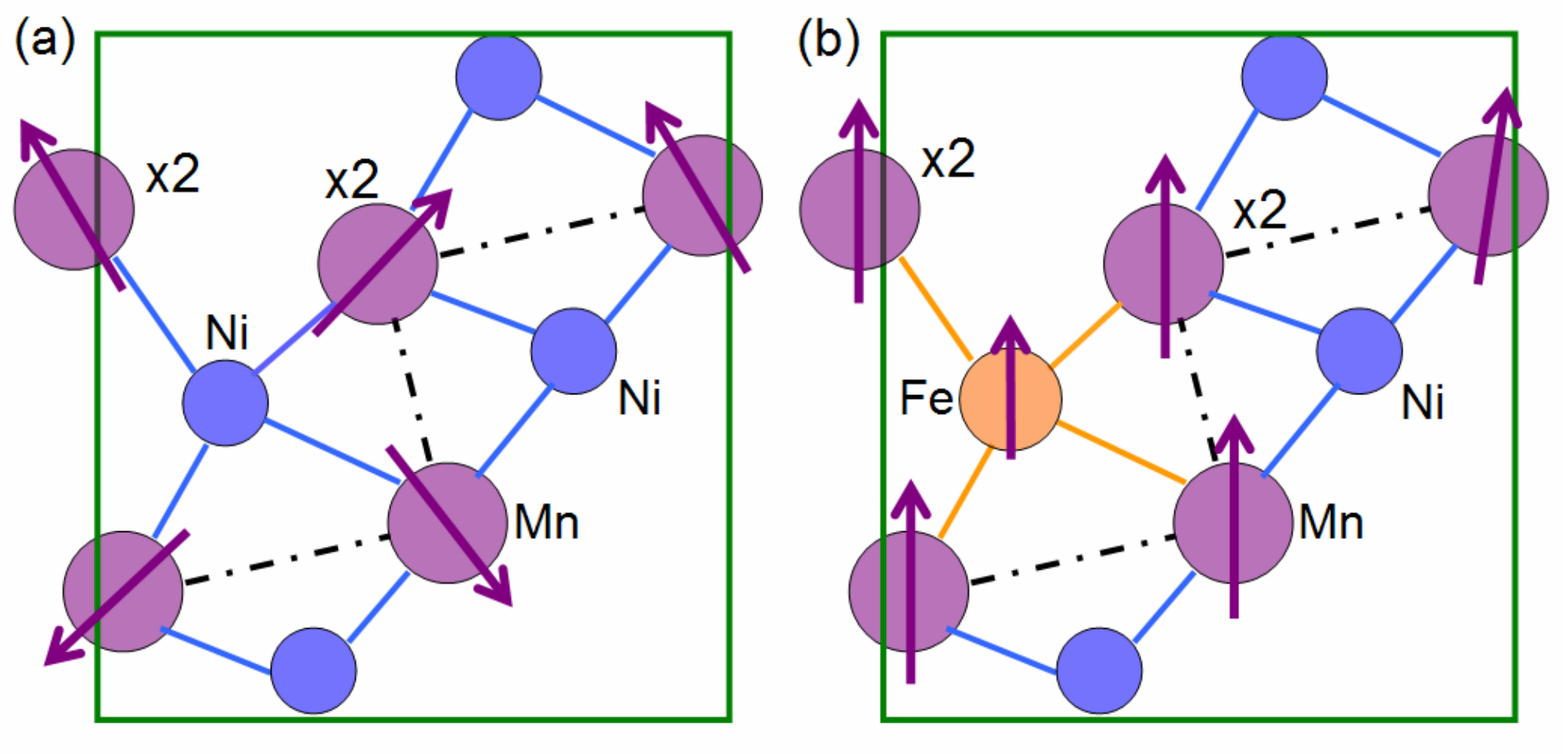

Fig. 4

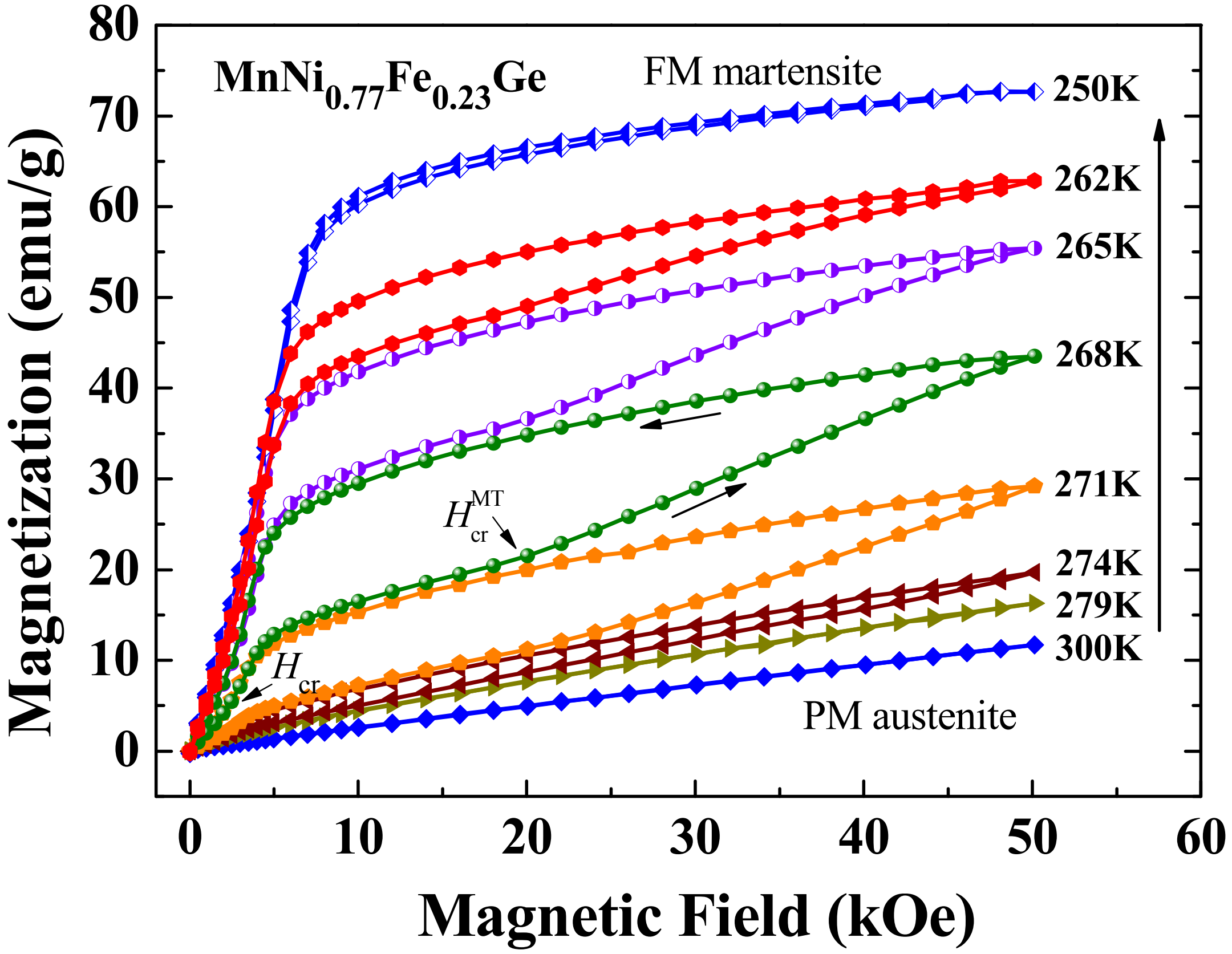

Fig. 5